\documentclass[conference,compsoc]{IEEEtran}

\ifCLASSOPTIONcompsoc
  \usepackage[nocompress]{cite}
\else
  \usepackage{cite}
\fi

\usepackage[numbers]{natbib}
\usepackage[pdftex]{graphicx}
\usepackage{float} %
\usepackage{amsmath}
\usepackage{bm} %
\usepackage{tcolorbox}
\usepackage{multirow}
\usepackage{multicol}
\usepackage{amsthm}
\usepackage{url}
\usepackage{amssymb}
\usepackage[hidelinks]{hyperref} %
\usepackage[capitalize,noabbrev]{cleveref}

\usepackage{placeins}
\usepackage{bbm}
\usepackage{booktabs}
\usepackage{rotating}
\usepackage{subcaption}
\usepackage{pdfpages}

\makeatletter
\renewcommand\paragraph{%
  \@startsection {paragraph}
    {4}
    {\parindent}
    {0ex plus 0.1ex minus 0.1ex}
    {0ex}
    {\normalsize \bf}
}
\makeatother

\begin{document}

\title{Membership Inference Attacks on Sequence Models}
\author{
    \IEEEauthorblockN{
        Lorenzo Rossi\IEEEauthorrefmark{1},
        Michael Aerni\IEEEauthorrefmark{2},
        Jie Zhang\IEEEauthorrefmark{2},
        Florian Tramèr\IEEEauthorrefmark{2}
    }
    \IEEEauthorblockA{\IEEEauthorrefmark{1}CISPA Helmholtz Center for Information Security, Saarbrücken, Germany}
    \IEEEauthorblockA{\IEEEauthorrefmark{2}ETH Zurich, Zurich, Switzerland}
    \IEEEauthorblockA{
        \texttt{lorenzo.rossi@cispa.de,}\\
        \texttt{\{michael.aerni,jie.zhang,florian.tramer\}@inf.ethz.ch}
    }
}

\maketitle
\begin{abstract}
Sequence models, such as Large Language Models (LLMs) and autoregressive image generators, have a tendency to memorize and inadvertently leak sensitive information. While this tendency has critical legal implications, existing tools are insufficient to audit the resulting risks. We hypothesize that those tools' shortcomings are due to mismatched assumptions. Thus, we argue that effectively measuring privacy leakage in sequence models requires leveraging the correlations inherent in sequential generation. To illustrate this, we adapt a state-of-the-art membership inference attack to explicitly model within-sequence correlations, thereby demonstrating how a strong existing attack can be naturally extended to suit the structure of sequence models. Through a case study, we show that our adaptations consistently improve the effectiveness of memorization audits without introducing additional computational costs. Our work hence serves as an important stepping stone toward reliable memorization audits for large sequence models.
\end{abstract}
\IEEEpeerreviewmaketitle

\section{Introduction}

Large sequence models, such as LLMs, are trained on a vast portion of the Internet. Since this data contains sensitive or protected information~\citep{tramer2022position,brown2022does}, understanding and mitigating memorization of sequence models is an important task. Membership inference attacks (MIAs~\citep{shokri2017membership}) are a key tool for this task, which aims to classify whether a given sample was part of the training data. However, current MIAs for large sequence models cannot reliably prove the presence or absence of memorization~\citep{zhang2024position}.

\begin{figure}[tb]
\centering
\includegraphics[width=\linewidth]{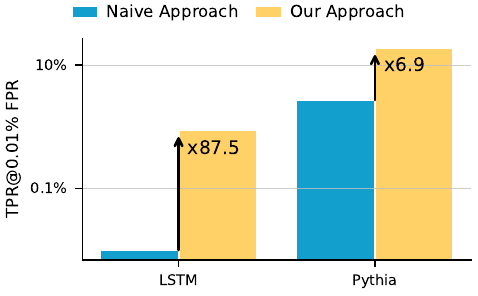}
\caption{\textbf{Strong membership inference for sequence models must consider correlations between sequence elements.} We apply LiRA, a state-of-the-art MIA, to measure memorization in language modeling. First, we use LiRA as-is, where membership guesses only depend on a sample's loss (``Naive Approach''). This approach uncovers only a small amount of memorization, as indicated by TPR (True Positive Rate) values close to the FPR (False Positive Rate) of 0.01\%. However, if we adapt LiRA to explicitly consider correlations between sequential predictions (``Our Approach''), we can uncover significantly more memorization (up to 87.5x more). See \cref{app:figure} for details.
}
\label{fig:teaser}
\end{figure}

In contrast, typical discriminative settings (e.g., image classification) exhibit highly successful membership inference attacks such as the Likelihood Ratio Attack (LiRA)~\citep{carlini2022membership}. Conceptually, LiRA first trains shadow models~\cite{shokri2017membership} that mimic a victim model and then uses a sample's cross-entropy loss on the victim and shadow models as the input to a statistical hypothesis test.\footnote{In practice, the loss is rescaled to better match statistical assumptions.} Hence, one might expect the same methodology to yield a strong baseline for large sequence models. However, a naive application can fail to uncover significant memorization, as seen in \cref{fig:teaser} (``Naive Approach'').

We hypothesize that the implicit assumptions of that approach are partially invalid for sequence models. In particular, standard MIAs assume that a sample's loss is sufficient to determine membership. However, in the context of sequence modeling, this treats all sequence elements as independent---an assumption that is clearly wrong for domains such as language. We hence argue that strong MIAs for sequence models crucially need to exploit the correlation between sequence elements.

In this paper, we explore adaptations to LiRA that account for within-sequence correlations. By explicitly estimating the covariance between sequence elements in a sample-efficient way, our adaptations uncover significantly more memorization in \cref{fig:teaser} than the naive approach---with only few shadow models and trivial overhead.

We then study our adaptations on different types of sequence models, where we show that those improvements are consistent: accounting for within-sequence correlations can significantly help an attack to uncover memorization, and never truly hurts. Additionally, we compare different types of covariance estimators, and find that a strong inductive bias can sometimes reduce the required number of shadow models by an order of magnitude. Therefore, while we do not propose a new standalone attack, our discoveries serve as an important stepping stone toward truly stronger MIAs for large sequence models.

\section{Preliminaries and Related Work}
We first describe memorization and membership in general, and then introduce autoregressive models as one specific instance of sequence models.

\subsection{Memorization and Privacy Auditing}
The extent of memorization in sequence models remains an active area of debate. Although some studies suggest that large-scale models exhibit limited memorization and are unlikely to expose sensitive information under normal conditions~\citep{anil2023palm,reid2024gemini,achiam2023gpt,duan2024membership}, other works provide strong evidence that these models can memorize and leak substantial portions of their training data~\citep{carlini2022privacy,carlini2022quantifying,carlini2021extracting,nasr2023scalable,pinto2024extracting,somepalli2023diffusion,291199,aerni2024measuring}. The discrepancy between these findings is probably due to differences in evaluation methodologies, dataset characteristics, and the selection of test samples (some types of data are more prone to memorization~\citep{carlini2022privacy,aerni2024evaluations}).

Beyond empirical evidence, theoretical studies indicate that memorization is not simply an unintended byproduct of large-scale training, but it can be even necessary to achieve a low generalization error~\citep{feldman2020does,wang2024memorization}. This raises fundamental questions about the trade-off between model utility and privacy risks. Understanding the mechanisms of memorization, identifying vulnerable data instances, and designing robust privacy auditing techniques are essential to mitigating the risks associated with sequence models.

\subsection{Membership Inference Attacks}
The goal of MIAs is to determine whether a specific sample was part of the training data~\citep{shokri2017membership}. This is typically framed as a standard security game~\citep{yeom2018privacy,jayaraman2020revisiting,carlini2022membership}, where the objective is to make a binary prediction—indicating whether the sample was included in the training dataset or not.
Rather than making strict binary predictions, MIAs often rely on a membership inference score $\mathcal{A}(f, x)$, where  $f$, $x$ are respectively the model and sample audited. A higher scores reflect greater confidence that a sample belongs to the training dataset. A threshold can always be applied to convert these scores back into binary decisions.
Using a soft prediction approach enables the evaluation of performance across all possible false positive rates. This, in turn, allows for a comprehensive analysis of the full ROC curve.

When MI is employed as a privacy auditing tool, as is frequently done in computer security~\citep{ho2017detecting,aerni2024evaluations,carlini2022membership,DPorg-average-case-dp}, these attacks are usually assessed using worst-case metrics such as the attack's true positive rate (TPR) at low false positive rates (FPR). Consequently, if a membership inference attack can reliably and consistently compromise the privacy of even a small subset of vulnerable users in a sensitive dataset, the training algorithm can be considered to leak private information (see \cite{carlini2022membership,aerni2024evaluations} for a more detailed discussion).

Following the standard membership inference security game is often impractical, as it would require training too many models to get reasonable estimates for the TPR @ low FPR metric. For this reason, it is common to sample a random subset of the canaries and add them to the training data at each run~\citep{carlini2022membership,aerni2024evaluations,steinke2024privacy}. %

Different membership inference attacks were developed specifically for LLMs~\citep{shi2023detecting,zhang2024min, mattern2023membership}. Some of the existing black-box attacks exploit the sequential structure of language and construct a statistic by aggregating the per-token scores in a specific way. For instance, \cite{shi2023detecting} showed that only considering the $K\%$ of tokens with the smallest likelihood improves the performance of the attack.

Many works use the same global threshold across all the samples. However, different studies have shown that calibrating the threshold for each sample greatly improves membership inference attacks, as some samples are harder to learn~\citep{carlini2022membership,sablayrolles2019white,watson2021importance}. This is particularly problematic because most of the uncalibrated membership inference attacks are strong \textit{non-membership inference attacks}; they cannot reliably distinguish \textit{easy-to-predict non-members} and \textit{hard-to-predict members}~\citep{watson2021importance,carlini2022membership}. Several studies address this issue by calibrating predictions~\citep{carlini2022membership,tramer2022truth,zarifzadeh24a} using numerous shadow models. While employing a large number of shadow models improves MIA performance, it also comes with a major drawback—significantly increasing computational costs. In the following, we focus on the Likelihood Ratio Attack as a state-of-the-art instance of such attacks.

\paragraph{The Likelihood Ratio Attack (LiRA)} \citep{carlini2022membership} developed a method to calibrate the scores using shadow models. The main idea is to train many shadow models with and without the target sample $x$. Then, compute a score $S$ for each run, and estimate the parameters of a Gaussian distribution for the IN case, where the target sample $x$ was part of the training data, and the OUT case, where it was not part of the training data. Finally, compute the likelihood ratio%
\begin{equation*} %
    \mathcal{A}(f, x) := \frac{\mathcal{N}(S(f; x) \mid \mu_{x,\text{in}}, \sigma_{x,\text{in}}^2)}{\mathcal{N}(S(f; x) \mid \mu_{x,\text{out}}, \sigma_{x,\text{out}}^2)} \, .
\end{equation*}

They also noticed that using scores of multiple augmented versions of the same sample improves the performance. To account for multiple scores, they model each IN and OUT score as an independent Gaussian distribution.

The Neyman-Pearson lemma~\citep{neyman1933ix} ensures that the likelihood ratio test is the most powerful test for a given model. In LiRA, this model assumes a Gaussian loss distribution, making the test uniformly most powerful under that assumption. If the true loss distribution deviates from Gaussianity, the resulting likelihood ratio test may be suboptimal in practice, limiting the expressiveness and effectiveness of the attack.

\subsection{Autoregressive Sequence Models}
Autoregressive models are a common way to solve a wide range of problems across various sequential domains, including text generation~\citep{larochelle2011neural}, image generation~\citep{el-nouby2024scalable,salimans2017pixelcnn}, time series forecasting~\citep{box1968some}, and protein structure prediction~\citep{shin2021protein}. The underlying assumption of autoregressive models is that each sequence element only depends on its preceding elements, formalized as follows:
\begin{align*}
    x_{t+1} \sim p(x_{t+1} \mid x_1, ...,x_t), \quad t \in \{1,2,\dots\} \, ,
\end{align*}
where the conditional probability distribution $p(x_{t+1} \mid x_1, ...,x_t)$ models the likelihood of the next element given the past observations. The likelihood of a given sequence is hence the product of individual elements' likelihoods.

Several architectures have been developed for different types of sequential data:

\paragraph{Recurrent Neural Networks (RNNs) and Long Short-Term Memory (LSTM)} Recurrent neural networks (RNNs), particularly Long Short-Term Memory (LSTM) networks~\citep{hochreiter1997long}, have historically been used due to their ability to capture long-range dependencies.

\paragraph{Transformer-based architectures and LLMs} Transformer architectures~\citep{vaswani2017attention} revolutionized autoregressive modeling, particularly in text generation. Unlike RNNs and LSTMs, transformers rely entirely on self-attention mechanisms, allowing them to capture long-range dependencies more effectively. LLMs, such as Pythia~\citep{biderman2023pythia}, build on this foundation, scaling transformers to billions of parameters to achieve human-like text generation capabilities.

\paragraph{Autoregressive Image Generation} Autoregressive models have been extended to image generation by sequentially modeling the dependencies between pixels. For example, PixelCNN++~\citep{salimans2017pixelcnn} employs a convolutional architecture that efficiently captures these pixel dependencies while preserving the autoregressive property. %

\subsection{Covariance Estimation}
LiRA models a sample's loss using Gaussian distributions. For autoregressive models, a sample's loss is the sum of per-element losses. An alternative is hence to model the losses of individual sequence elements as a multivariate Gaussian, which requires estimating a covariance.

The simplest approach to estimating the covariance matrix is through the maximum likelihood estimator (MLE):
\begin{align*}
    \Sigma_{\text{MLE}} = \frac{1}{N} \sum_{i=1}^{N} (X_i - \bar{X})(X_i - \bar{X})^T ,
\end{align*}
where the $\{X_i\}_{i=1}^{N}$ are the $N$ observed vectors, and $\bar{X} =\frac{1}{N} \sum_{i=1}^{N} X_i$. In LiRA, $N$ represents the number of shadow models used, and $X_i$ represents the vector of (log) losses of the $i$-th shadow model. Although the MLE is unbiased, it has notable shortcomings. In particular, it often fails to provide reliable estimates of the eigenvalues of the covariance matrix, and in scenarios where the number of samples is smaller than the number of dimensions, the resulting matrix is non-invertible. This poses significant challenges, especially in high-dimensional applications where an accurate MLE requires many samples.

One straightforward solution is to assume a diagonal covariance structure, which effectively treats different features as uncorrelated. A more advanced type of structure is given by shrinkage-based covariance estimators. These methods compute a convex combination of the MLE-based covariance matrix and a diagonal matrix with controlled variance. The key idea is to balance the variance of the sample covariance with the bias introduced by the target matrix, thereby improving the overall estimation quality.

A particularly effective shrinkage estimator is the Oracle Approximating Shrinkage Estimator (OAS)~\citep{chen2010shrinkage}. OAS improves the estimation of covariance matrices by adaptively selecting the shrinkage intensity based on the characteristics of the sample data. Rather than using a fixed shrinkage parameter, OAS computes the optimal intensity that minimizes the expected error (typically measured by the mean squared error) between the true covariance matrix and the shrinkage estimator. In practice, the OAS estimator constructs the covariance estimate as follows:
\begin{align}
\Sigma_{\text{OAS}} = (1 - \alpha) \Sigma_{\text{MLE}} + \alpha F,    
\end{align}
where $\Sigma_{\text{MLE}}$ is the sample covariance matrix estimated via MLE, $F$ is $\frac{\operatorname{tr}(S)}{p} I$, $p$ is the dimensionality of the data, $I$ is the identity matrix, and $\alpha$ is the shrinkage intensity determined from the data.

\section{Membership Inference on Sequence Models}  \label{sec:mia-seq}
In the following, we focus on language models for simplicity and hence use the terms token and sequence elements interchangeably. However, all points apply for more general sequence models, such as autoregressive image generators (see \cref{app:image} for an example). Moreover, we use LiRA as a baseline MIA, but our observations also apply to other shadow model-based attacks such as \cite{zarifzadeh24a,ye2022enhanced}.

\subsection{Mismatched Assumptions in Naive LiRA}
We hypothesize that the standard LiRA attack has implicit assumptions that can be invalid for sequence models. Concretely, those assumptions are
\begin{enumerate}
    \item \textbf{Independence}: The membership signal is sufficiently captured by a sample's loss. Multiple statistics for a single sample are independent.
    \item \textbf{Heteroscedasticity}: the variance of a sample's loss (or a derived statistic) depends on whether the sample is in the training data or not.
\end{enumerate}
In the following, we discuss how those assumptions might be invalid for sequence models and propose adaptations.

\paragraph{Independence}
For traditional classification models, such as image classifiers, a sample's cross-entropy loss depends on the entire sample. In contrast, autoregressive models typically calculate a loss per token, such that a sample's loss is the sum (or mean) of per-token losses.
Hence, a naive application of LiRA---which only uses the loss of the entire sequence---corresponds to modeling the token losses as independent Gaussians.

However, the per-token losses of autoregressive models are highly dependent by definition! We hence model the entire per-token loss vector as a multivariate Gaussian, thereby explicitly accounting for inter-token correlations. This results in the following modified hypothesis test, where the IN and OUT distributions are now multivariate Gaussians:
\begin{equation*} %
    \mathcal{A}(f, x)
    := \frac{
    \mathcal{N}(\mathbf{S}(f; x) \mid {\mathbf{\mu}}_{x,\text{in}}, \mathbf{\Sigma}_{x,\text{in}})
    }{
    \mathcal{N}(\mathbf{S}(f; x) \mid {\mathbf{\mu}}_{x,\text{out}}, \mathbf{\Sigma}_{x,\text{out}})
    } \, ,
\end{equation*}
where the score function $\mathbf{S}(\cdot, \cdot)$ applies element-wise for every per-token loss, ${\mathbf{\mu}}_{x,\text{in}}, {\mathbf{\mu}}_{x,\text{out}}$ are mean vectors with dimensionality equal to the sequence length $T$, and $\mathbf{\Sigma}_{x,\text{in}}, \mathbf{\Sigma}_{x,\text{out}}$ are $T \times T$ covariance matrices.

\begin{figure}[tp]
\centering
\includegraphics[width=\linewidth]{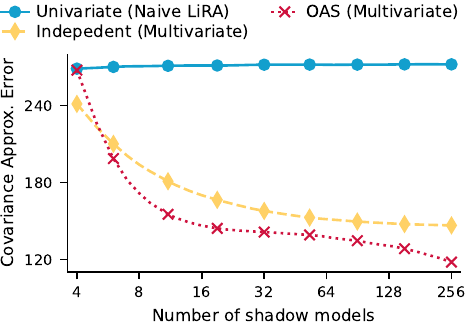}
\caption{\textbf{Assuming independence between per-token losses can yield a large approximation error.}
We use the full covariance MLE (with 484 shadow models) as the gold standard,
and measure the distance in Frobenius norm (``Covariance Approximation Error'')
to other models' estimates.
As the number of shadow models increases, we analyze this error using 1,000 average-case canaries from the IN case of Pythia 1b with a sequence length of 128. See \Cref{fig:cov_0} in \Cref{app:covariance_0} for the corresponding figure for the OUT case, and for more details, see \Cref{app:figure}.}
\label{fig:full_covariance-independence}
\end{figure}

We illustrate this model mismatch in \cref{fig:full_covariance-independence} for sequences of 128 tokens. First, we estimate the full covariance matrix using many shadow models (484) for an LLM (Pythia 1b). We then compute the covariance matrices according to the original LiRA model ($\mathbf{\Sigma} = \sigma^2 \mathbf{I}$; ``Naive LiRA (Univariate)'') and a model that estimates per-token loss variances but no correlations ($\mathbf{\Sigma} = \text{diag}(\sigma_1^2, \dotsc, \sigma_T^2)$; ``Independent (Multivariate)'').\footnote{We drop the in/out suffix for brevity.} Finally, we compute the distance of the full covariance matrix to the two approximations in terms of Frobenius norm. Both approximations do not converge to the full covariance, and have a large approximation error---especially for few shadow models.

Unfortunately, the full covariance matrix has a parameter count that is quadratic in the number of tokens. This significantly increases the sample complexity required for accurate covariance estimation. Because the number of samples corresponds to the number of shadow models, the larger sample complexity also increases computational cost.

To obtain the richer model with a manageable computational cost, we explore efficient covariance estimation techniques. The OAS estimator~\citep{chen2010shrinkage} is particularly well-suited for high-dimensional settings with limited samples. As seen in \cref{fig:full_covariance-independence} (``OAS (Multivariate)''), OAS approximates the full covariance more accurately with few shadow models, and eventually matches the full covariance.

\begin{figure}[tp]
\centering
\includegraphics[width=\linewidth]{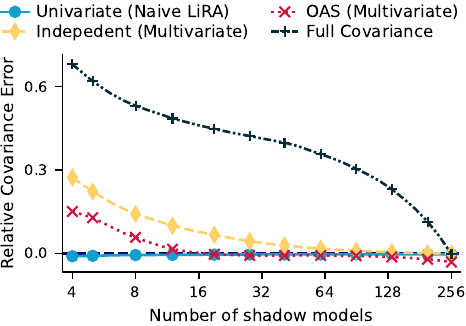}
\caption{\textbf{Assuming shared covariances (homoscedascity) is empirically beneficial.}
The Relative Covariance Error is measured as the relative difference between the Covariance Approximation Error of the Class-Wise and Shared covariance estimates. As the number of shadow models increases, we analyze this error using 1,000 average-case canaries from the IN case of Pythia 1b with a sequence length of 128. See \Cref{fig:diff_cov_0} in \Cref{app:covariance_0} for the corresponding figure for the OUT case, and for the experimental details, see \Cref{app:figure}.}
\label{fig:full_covariance-heteroscedasticity}
\end{figure}

\paragraph{Heteroscedasticity} Standard LiRA assumes that the covariance of losses differs for in vs.\ out distributions (heteroscedasticity). However, for sequence models, we find that using the same covariance matrix for both distributions (homoscedascity) is empirically more accurate. This effectively doubles the number of available samples for covariance estimation, as we use the shadow models from both the in and out distributions to estimate the covariance matrix.

Concretely, we use 484 shadow models to obtain an accurate estimate of the full covariance matrices to serve as a gold standard. We then estimate covariances with different structures for a varying number of shadow models, once class-wise (homoscedastic), and once shared between classes (heteroscedastic). We then calculate the approximation error in terms of Frobenius norm between the estimates and the gold standard, and plot the relative difference (class-wise error minus shared error, normalized by the shared error). A large relative difference indicates that the class-wise covariance estimate has a much larger approximation error than the shared covariances.

\Cref{fig:full_covariance-heteroscedasticity} illustrates the benefits of using the same in and out distributions.
We find that using the same covariance for in and out distributions is typically beneficial, as indicated by the mostly positive relative difference. Those benefits are particularly pronounced when the number of shadow models is small; we conjecture that this is again due to the reduction in sample complexity.

\subsection{Attack Variants}
Based on the previous observations, we consider the following concrete adaptations for LiRA on sequence models:

\begin{itemize}  
    \item \textbf{Univariate:} A baseline approach where per-token scores are averaged into a single scalar, and a univariate Gaussian distribution is estimated for each class (member and nonmember). This corresponds to a naive application of LiRA to sequence models.
    \item \textbf{Independent:} Uses per-token scores directly but retains the independence assumption, estimating only the diagonal tokens of the covariance matrix.  
    \item \textbf{OAS:} Drops the independence assumption by estimating the covariance matrix using OAS. 
\end{itemize}  

Additionally, based on the Heteroscedasticity assumption,  each of these methods can be further divided:
    \begin{itemize}  
        \item \textbf{Class-Wise:} Estimates separate covariance matrices for members and nonmembers.  
        \item \textbf{Shared:} Uses a single covariance matrix for both classes, leveraging all available samples for estimation. In this case, we first center the members and nonmembers using per-class means, but then estimate a shared covariance matrix.
    \end{itemize}

In some cases, tokens in the sequence may be redundant or introduce unnecessary noise. To address this, we explore methods to reduce the sequence length while preserving essential information:
\begin{itemize}
    \item \textbf{Group.} The sequence is divided into fixed-size chunks, and the average score is computed for each chunk. This approach smooths local variations while retaining overall trends.
    \item \textbf{Min.} The token with the smallest score is selected. This follows a similar intuition as Min-K\%, emphasizing the most confident (least anomalous) tokens in the sequence.
    \item \textbf{Max.} Analogous to Min, but instead selects the token with the largest score, capturing the most uncertain or anomalous regions in the sequence.
\end{itemize}

\section{Experiments} \label{sec:experiments}

\subsection{Setup} \label{sec:settings}
We study attack variants across two language modeling tasks: training an LSTM from scratch and fine-tuning a transformer-based LLM. This comparison provides insights into how different generative models memorize and expose sensitive information under various conditions. Furthermore, in \cref{app:image}, we describe a case study using Pixel-CNN++~\citep{salimans2017pixelcnn}, an autoregressive image generator. Our objective is to assess the performance of membership inference attacks (MIA) under different model architectures and training paradigms.

We use two types of canaries: average-case and worst-case. We sample average-case canaries uniformly from the test set. They hence represent naturally occurring sequences within the dataset and help measure memorization under typical conditions. For worst-case canaries, we aim to pick samples that are particularly prone to memorization. Following common wisdom~\citep{anil2023palm,team2023gemini}, we use random token sequences as worst-case canaries. These synthetic sequences serve as adversarial inputs designed to maximize memorization effects, thereby approximating an upper bound of privacy leakage. In both cases, we consider 10,000 canaries.

We train all the language models on the PersonaChat dataset~\citep{zhang2018personalizing}, which consists of conversations of people describing themselves. This dataset mimics a realistic setting where privacy leakage could be a concern. For each architecture and canary type, we train a total of 64 models, ensuring that each canary appears in exactly half of the model's training data. We then adopt a common leave-one-out strategy for evaluation. That is, we treat each of the 64 models as a target and use the remaining 63 models as shadow models.

\paragraph{LSTM-Based Models} \label{subsec:lstm setting}  
We train the LSTM models~\citep{hochreiter1997long} from scratch using a standard architecture. The architecture consists of an initial embedding layer to encode tokens, followed by two LSTM layers with a hidden dimension of 192 and a fully connected prediction head for token classification. We use the Pythia 1b tokenizer and train the model on the standard next-token prediction task with cross-entropy loss. See \Cref{tab:hyperparameters} from \Cref{app:hyperparameters} for the specific choice of hyperparameters.

\paragraph{Transformer-Based Models (Pythia 1b)} \label{subsec:pythia setting}  
For both average- and worst-case canaries, we fine-tune the deduplicated Pythia 1b model~\citep{biderman2023pythia}, a 1-billion-parameter transformer-based language model pre-trained on a large corpus of data from the internet. The training follows the standard next-token prediction objective using cross-entropy loss. To further explore the importance of the number of shadow models, we fine-tune a large number (484) of shadow models using average-case canaries. As in the other settings, we run leave-one-out on all the 484 shadow models. See \Cref{tab:hyperparameters} from \Cref{app:hyperparameters} for specific hyperparameters.

\paragraph{Evaluation}
We focus the evaluation on the true positive rate (TPR) in the low false positive rate (FPR) regime, specifically on TPR@0.01\% FPR. 
This metric identifies instances of memorization with high confidence, aligning with standard practices in privacy auditing (see \citep{carlini2022membership,aerni2024evaluations} for a detailed discussion).

\subsection{Comparison of Different Attacks}\label{subsec:attacks}
\begin{figure}[!t]
\centering
\includegraphics[width=\linewidth]{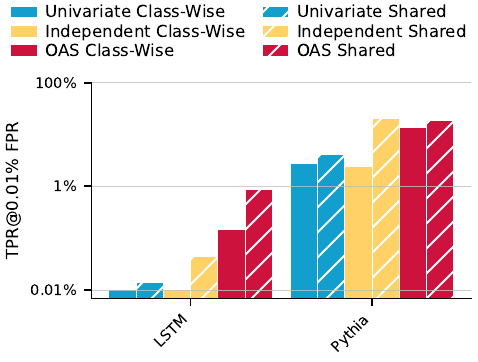}
\caption{\textbf{The Univariate Class-Wise approach (Naive LiRA) results in a weaker attack.} TPR@0.01\% FPR for various attacks using 10,000 average-case canaries with 64 shadow models. The Shared version consistently achieves better performance. Refer to \Cref{tab:results}, in \Cref{app:details}, for MIA performance with worst-case canaries.
}
\label{fig:bar_0}
\end{figure}

We first present an overview of LiRA adaptations using the different covariance estimators and a class-wise vs.\ shared covariance. Concretely, we compare all six attack variants using 64 shadow models with average-case canaries in \cref{fig:bar_0}. Additionally, in \cref{app:details}, \Cref{tab:results} shows the results for the worst-case canaries and the image domain. Moreover, \Cref{app:roc} highlights the Receiver Operating Characteristic (ROC) curve for the LSTM and Pythia average-case canaries, and shows that in these settings, the OAS systematically matches or beats the Univariate and Independent approaches across the whole curve.

First, compared to a naive baseline (``Univariate Class-Wise''), adaptations for sequence models seem to consistently uncover an order of magnitude more memorization in most cases. In particular, assuming homoscedasticity (``Shared'') alone can significantly increase the TPR at a low FPR and never hurts. 
When considering worst-case canary, the models memorize significantly more, and in particular, the Independent Shared approach has a TPR @ 0.01\% greater than 95\% for both LSTM and Pythia 1b.
In \Cref{app:length-reduction}, we examine various length reduction strategies and show that discarding certain tokens can enhance performance. For example, \Cref{fig:group-lstm-average} illustrates that the ``Min'' strategy—retaining only the two tokens with the smallest loss—yields a higher true positive rate (TPR).

The optimal attack method and length reduction strategy vary depending on the specific setting. In particular, it is important to analyze the covariance matrix to understand which prior information is useful in that specific setting. For instance, if the real covariance matrix is close to a diagonal matrix, and there is limited interaction between the tokens, then using the Independent approach leads closer to the real covariance matrix with a smaller number of shadow models than using the OAS method. Vice versa, if the real covariance matrix has many interactions the OAS approach becomes a more suitable choice.

\subsection{Impact of the Number of Shadow Models}
\begin{figure}[t]
\centering
\includegraphics[width=\linewidth]{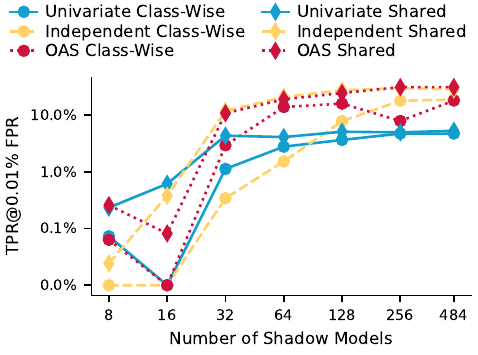} %

\caption{\textbf{The Univariate Shared attack is slightly better for small numbers of shadow models ($<16$), but it is significantly worse for large numbers of shadow models.} The reported TPR@0.01\% FPR results illustrate the performance of various attacks as the number of shadow models increases, based on experiments conducted with Pythia 1b and 10,000 average-case canaries.
}
\label{fig:shadow_models}
\end{figure}

The number of shadow models plays a crucial role in calibrating the attack, as it directly influences membership inference performance. To better understand this relationship, we analyze how attack effectiveness changes as a function of the number of shadow models, using a setting where we have access to a larger pool of 484 shadow models. 
As illustrated in \cref{fig:shadow_models}, increasing the number of shadow models improves attack performance by uncovering more memorization, leading to higher TPR at low FPR. However, we observe key differences across methods. The univariate attacks plateau early, peaking at 32 shadow models, while more expressive estimators continue improving with additional models. Notably, for small numbers of shadow models ($\leq 16$), the Univariate Shared attack slightly outperforms others, suggesting that a simpler approach yields higher TPR in this setting. Additionally, shared covariance estimators consistently achieve comparable or superior TPRs with fewer shadow models than the class-wise approach, making them a more efficient choice when computational resources are limited.

\section{Conclusion}
In this work, we question the assumptions of standard membership inference attacks in the context of large sequence models. Based on the resulting insights, we introduce adaptations for existing membership inference attacks. Our results demonstrate that leveraging correlations between per-token losses, rather than relying solely on the average loss, significantly boosts attack performance. Most importantly, our adaptations enable strong audits with only few shadow models---a crucial requirement given the computational cost of training large sequence models. We hence provide an important stepping stone towards reliable memorization audits for large sequence models.

\bibliographystyle{IEEEtran}
\bibliography{bib}

\appendix
\renewcommand\thesubsectiondis{\thesubsection}
\subsection{Figure Setup}\label{app:figure}
Here, we describe the setup for the most relevant figures.

\paragraph{\Cref{fig:teaser}} The two groups of bars refer to the LSTM, and the Pythia 1b, which are the two language settings analyzed, using average-case canaries. As described in \Cref{sec:settings}, we used 10,000 samples and 64 shadow models. The "Naive Approach", which corresponds to LiRA as-is, where membership guesses only depend on a sample’s loss, corresponds to the Univariate Class-Wise approach, while "Our Approach" corresponds to OAS Shared.

\paragraph{\Cref{fig:full_covariance-independence} and \Cref{fig:full_covariance-heteroscedasticity}} The figures focus on the case with Pythia 1b and average-case canaries because we have access to a larger number of shadow models. To obtain the most reliable estimation of the full covariance matrix, we use all 484 shadow models and apply the MLE, which serves as gold standard. Our analysis is based on 1,000 samples from the average-case canaries.
In \Cref{fig:full_covariance-independence}, the y-axis represents the covariance approximation error, measured as the Frobenius norm between the gold standard estimate and the covariance matrix obtained using different methods with varying numbers of shadow models. We choose the Frobenius norm as it is one of the most commonly used norms for matrices.
For \Cref{fig:full_covariance-heteroscedasticity}, the y-axis represents the relative covariance error, which quantifies the relative difference between the covariance approximation error of the shared estimated covariance matrix and that of the class-wise estimated covariance matrix (class-wise error minus shared error, normalized by the shared error). A higher relative covariance error indicates that the shared estimation is better (a smaller covariance approximation error) than the class-wise one.

\subsection{Additional Covariance Estimations}\label{app:covariance_0}
\begin{figure*}
    \centering
    \begin{subfigure}[t]{0.49\textwidth}
    \includegraphics[width=\linewidth]{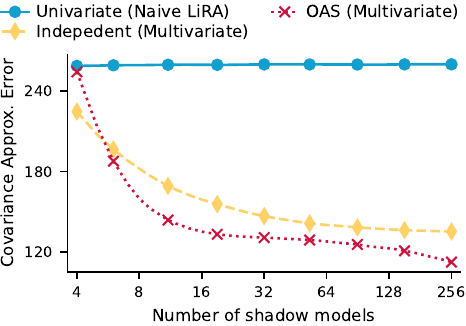}
    \caption{The Covariance Approximation Error, measured as the Frobenius norm between the covariance matrix estimated using 484 shadow models and the full covariance MLE, is compared to the covariance matrix estimated with a given method and varying numbers of shadow models. As the number of shadow models increases, we analyze this error using 1,000 average-case canaries from the OUT case of Pythia 1b with a sequence length of 128.}
    \label{fig:cov_0}
    \end{subfigure}
    \centering
    \begin{subfigure}[t]{0.49\textwidth}
    \includegraphics[width=\linewidth]{figures/diff_cov_average_1_text_7.pdf}
    \caption{The Relative Covariance Error is measured as the relative difference between the Covariance Approximation Error of the Shared estimated covariance matrix and the Class-Wise one. As the number of shadow models increases, we analyze this error using 1,000 average-case canaries from the IN case of Pythia 1b with a sequence length of 128.}
    \label{fig:diff_cov_0}
    \end{subfigure}
    \caption{\textbf{The OUT case follows the same trends as the IN case.}}
    \label{fig:both_cov_0}
\end{figure*}

\Cref{fig:cov_0} and \Cref{fig:diff_cov_0} present the corresponding results using the OUT case samples, analogous to \Cref{fig:full_covariance-independence} and \Cref{fig:full_covariance-heteroscedasticity}, respectively. The OUT case (distribution of the nonmembers' losses) shows the same trends as the IN case (distribution of the members' losses).

\subsection{Case Study: Autoregressive Image Generator}\label{app:image}
We also consider a case study using Pixel-CNN++~\citep{salimans2017pixelcnn}, an autoregressive image generator.

\textbf{Experimental Settings.}
We train 64 instances of Pixel-CNN++~\citep{salimans2017pixelcnn} on CIFAR-10. Due to the computational cost, the models are trained on a smaller number of epochs, compared to the original implementation which trained the model for 5000 epochs, the model is slightly under-trained leading to poor membership inference performance. Therefore, this shows an extremely challenging setting. We use samples from the test set as canaries. Compared to the language setting, where the sequence length is 128, in the image case the sequence length depends on the size of the image which is 32 x 32, and therefore the sequence length is 1024. This further increases the computational cost and the complexity of the task.

\begin{figure}[t]
    \centering
    \includegraphics[width=\linewidth]{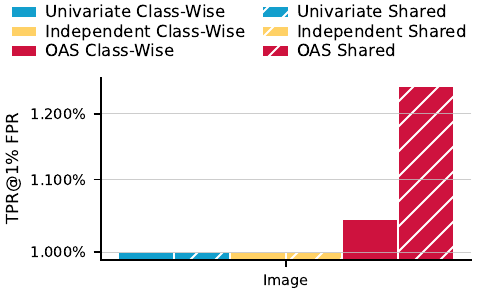}
    \caption{\textbf{OAS is the only approach that performs better than random guessing.} The TPR @ 1\% FPR using PixelCNN++ for average-case canaries.}
    \label{fig:main_results_2}
\end{figure}

\begin{figure}[t]
    \centering
    \includegraphics[width=\linewidth]{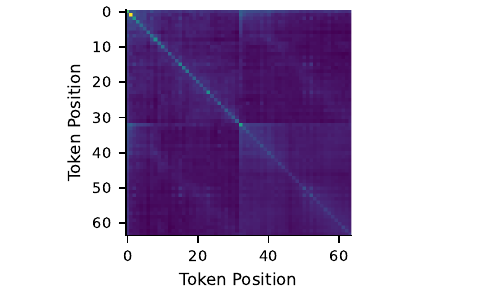}
    \caption{\textbf{The covariance structure is not diagonal.} The covariance matrix of the OUT case is estimated using the full covariance MLE on the distribution of the first 64 tokens with PixelCNN++ using 64 shadow models.}
    \label{fig:cov_image}
\end{figure}

\textbf{Attack Evaluation.} \Cref{fig:main_results_2} shows the performance of the MIAs with an Autoregressive Image Generator. Overall, the performance is quite poor. For this reason, we highlight the MIA performance using TPR @ 1\% FPR instead of TPR @ 0.01 FPR. In \Cref{app:details}, \Cref{tab:results} confirms that the performance for 0.1\% and 0.01\% is close to random guessing. OAS is the only attack that performs better than random guessing for TPR @ 1\% FPR. The reason is that there is a stronger interaction between tokens than in the language domain. To illustrate this, \Cref{fig:cov_image} shows the covariance matrix of the per-token loss distribution of the members. We observe that the structure is more complex and is different from the Independent one, as the covariance matrix has nonzero values outside the diagonal.

\subsection{Choice of the hyperparameters}\label{app:hyperparameters}
\Cref{tab:hyperparameters} shows the selected hyperparameters for each setting. In all the settings, we trained the models using AdamW~\citep{loshchilov2017decoupled}.

\begin{table*}
    \centering
    \begin{tabular}{l|cccccc}
\hline
Setting & Epochs & Learning Rate & Weight decay &  Batch Size & Hidden Dimension & Sequence Length \\
\hline
LSTM & 10 & $10^{-3}$ & 0.0 & 64 & 192 & 128 \\
Pythia 1b & 7 & $10^{-4}$ & $10^{-4}$ & 16 & / & 128 \\
Pixel-CNN++ & 50 & $10^{-3}$ & 0.0 & 192 & 192 & 1024 \\
\hline
    \end{tabular}
    \caption{\textbf{The selected hyperparameters for each setting}}
    \label{tab:hyperparameters}
\end{table*}

\subsection{ROC Curve}\label{app:roc}
\begin{figure*}[!htb]
\centering
\includegraphics[width=\textwidth]{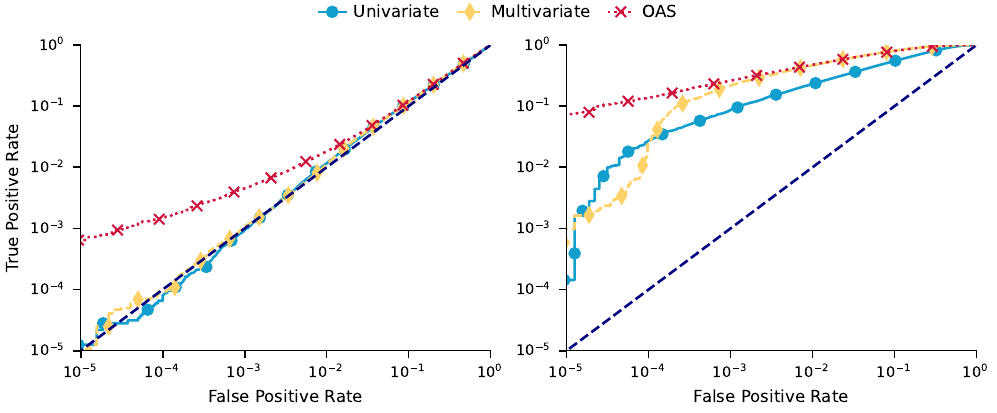}
\caption{\textbf{Using the Naive Approach (Univariate) gives suboptimal results.} Comparing the true positive rate vs. false positive rate for different ways to model the LiRA IN and OUT distributions  in the class-wise case. On the left, we use an LSTM, and on the right Pythia 1b using average-case canaries.}
\label{fig:roc_classwise}
\end{figure*}

\begin{figure*}[!htb]
\centering
\includegraphics[width=\textwidth]{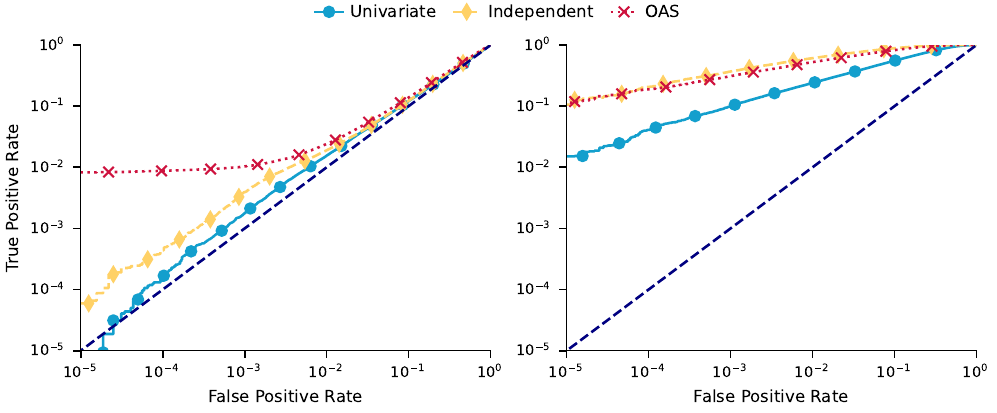}
\caption{\textbf{Using Naive Approach (Univariate) gives suboptimal results, also when considering a shared covariance matrix.} Comparing the true positive rate vs. false positive rate for different ways to model the LiRA IN and OUT distributions in the shared case. On the left, we use an LSTM, and on the right Pythia 1b using average-case canaries.}
\label{fig:roc_shared}
\end{figure*}

\Cref{fig:roc_classwise} and \Cref{fig:roc_shared} show the log-scale Receiver Operating Characteristic (ROC) curve of the success rates of the different types of attacks for the LSTM and Pythia Average. In particular, \Cref{fig:roc_classwise} shows the ROC curve using the two covariances (Class-Wise), while the \Cref{fig:roc_shared} shows the results with a single shared covariance matrix (Shared). In both cases, the results are for the average-case canaries. We see that the OAS approach has a higher TPR compared to the other in particular, for low FPR, where privacy auditing is more important.

\subsection{Detailed Results}\label{app:details}
\Cref{tab:results} shows the results for each setting. First, we observe that the sequence-aware attacks (Independent and OAS) obtain higher TPRs across all the settings. For the image case, all the attacks have a close to random guessing performance due to the complexity of the task.

\begin{table*}[tp]
\centering

\begin{tabular}{l|cccccccccc}
\hline
    & \multicolumn{2}{c}{Pythia Average} & \multicolumn{2}{c}{Pythia Worst} & \multicolumn{2}{c}{LSTM Average} & \multicolumn{2}{c}{LSTM Worst} & \multicolumn{2}{c}{Image}\\
MIA & 0.1\% & 0.01\% & 0.1\% & 0.01\% & 0.1\% & 0.01\% & 0.1\% & 0.01\% & 0.1\% & 0.01\%\\ \hline
 Univariate Class-Wise  & 8.69  & 2.72  & 95.04  & 89.66  & 0.10  & 0.01  & 92.51  & 4.33  & 0.10  & 0.01 \\
 Univariate Shared  & 10.01  & 4.09  & 94.91  & 91.26  & 0.18  & 0.01  & 95.82  & 88.25  & 0.10  & 0.01 \\ 
 \hline
 Independent Class-Wise  & 21.78  & 2.37  & 99.58  & 99.18  & 0.11  & 0.01  & 91.52  & 0.02  & 0.10  & \textbf{0.02} \\
 Independent Shared  & \textbf{36.98}  & \textbf{20.72}  & 99.67  & 99.30  & 0.39  & 0.04  & \textbf{98.75}  & \textbf{95.00}  & 0.10  & 0.01 \\
 OAS Class-Wise  & 26.26  & 13.81  & \textbf{99.78}  & \textbf{99.57}  & 0.45  & 0.15  & 23.19  & 0.03  & \textbf{0.11}  & 0.01 \\
 OAS Shared  & 30.95  & 18.81  & 99.67  & 99.37  & \textbf{1.03}  & \textbf{0.87}  & 70.74  & 30.72  & 0.10  & 0.01 \\
\hline
\end{tabular}

\caption{\textbf{The univariate attacks always under-perform the sequence-aware attacks.} The TPR@\{0.1,0.01\}\% FPR for different types of attacks and settings. All the scores are computed using 10,000 canaries and 64 shadow models using leave-one-out.}
\label{tab:results}
\end{table*}

\subsection{Comparison of Length Reduction Strategies}\label{app:length-reduction}
\begin{figure*}[t]
\centering
\includegraphics[width=0.9\linewidth]{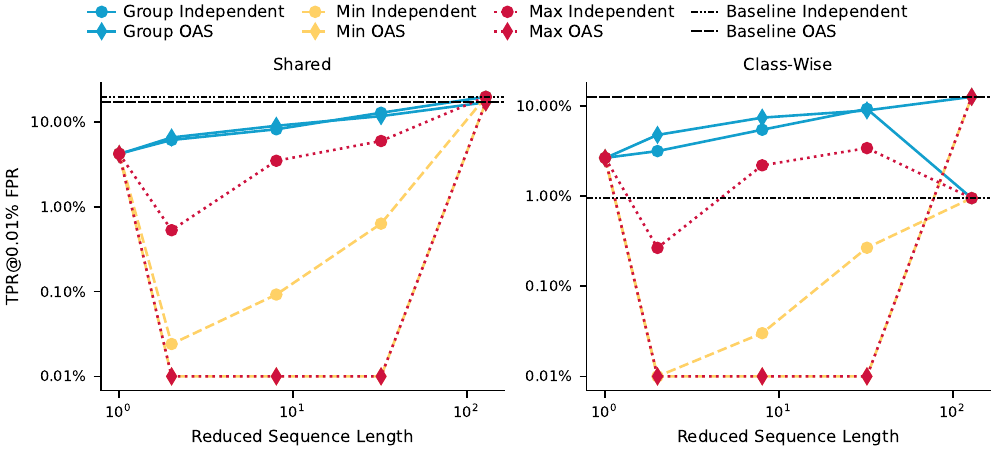}
\caption{\textbf{Pythia - Average-case canaries.} TPR @ 0.01\% FPR for different length reduction strategies on Pythia 1b with the average-case canaries. The black lines represent the baselines, where no grouping strategy is applied, and all the tokens in the sequence are used. The dashed lines represent the OAS case, while the solid lines represent the Independent case.}
\label{fig:group-pythia-average}
\end{figure*}

\begin{figure*}[th]
    \centering
    \includegraphics[width=\linewidth]{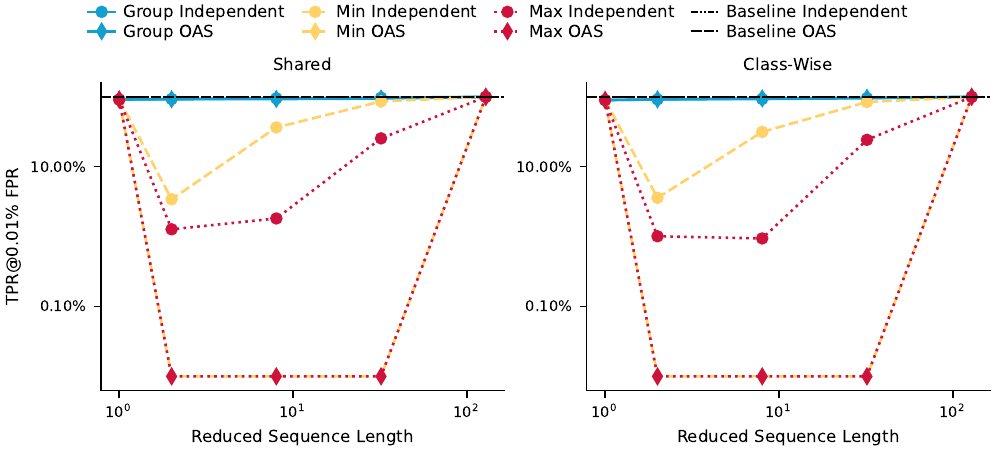}
    \caption{\textbf{Pythia - Worst-case canaries.} TPR @ 0.01\% FPR for different length reduction strategies on Pythia 1b with the worst-case canaries. The black lines represent the baselines, where no grouping strategy is applied, and all the tokens in the sequence are used. The dashed lines represent the OAS case, while the solid lines represent the Independent case.}
    \label{fig:group-pythia-worst}
\end{figure*}

\begin{figure*}[th]
    \centering
    \includegraphics[width=\linewidth]{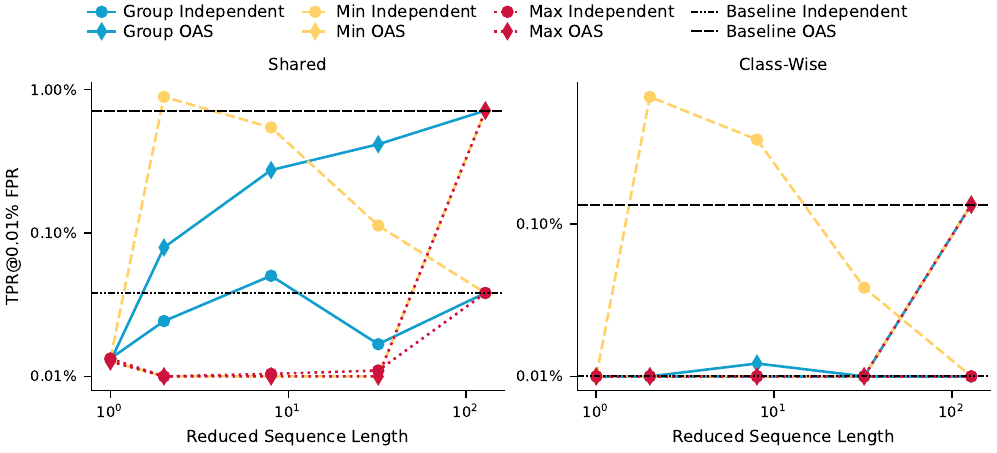}
    \caption{\textbf{LSTM - Average-case canaries.} TPR @ 0.01\% FPR for different length reduction strategies on LSTM with the average-case canaries. The black lines represent the baselines, where no grouping strategy is applied, and all the tokens in the sequence are used. The dashed lines represent the OAS case, while the solid lines represent the Independent case.}
    \label{fig:group-lstm-average}
\end{figure*}

\begin{figure*}[th]
    \centering
    \includegraphics[width=\linewidth]{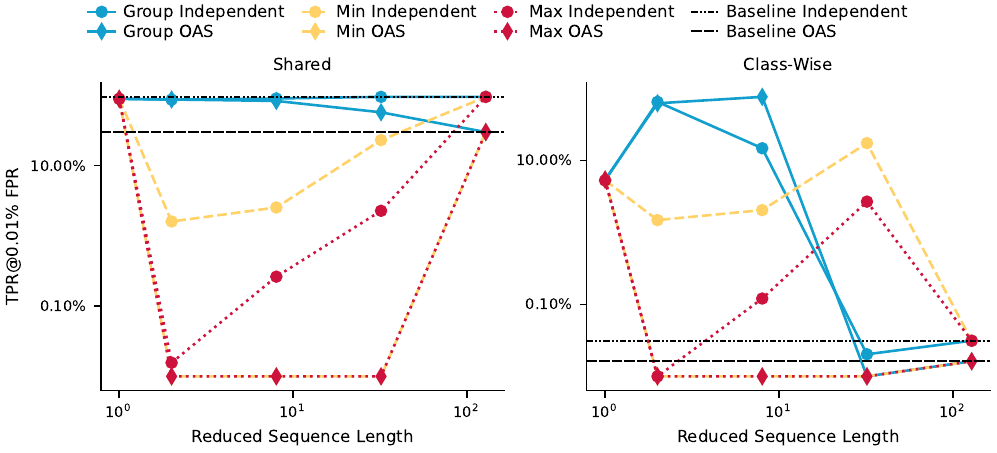}
    \caption{\textbf{LSTM - Worst-case canaries.} TPR @ 0.01\% FPR for different length reduction strategies on LSTM with the worst-case canaries. The black lines represent the baselines, where no grouping strategy is applied, and all the tokens in the sequence are used. The dashed lines represent the OAS case, while the solid lines represent the Independent case.}
    \label{fig:group-lstm-worst}
\end{figure*}

We further evaluate length reduction strategies that aim to select a representative subsequence for membership inference. Grouping corresponds to chunking the tokens in groups and computing the average of each group, therefore, when the sequence length is 1, this corresponds to Naive LiRA (Univariate Approach), and using all the 128 tokens corresponds to the standard baseline (either Independent or OAS).
\Cref{fig:group-pythia-average,fig:group-pythia-worst,fig:group-lstm-average,fig:group-lstm-worst} show the TPR @ 0.01\% FPR for different length reduction strategies.
For instance, \Cref{fig:group-pythia-average}, which represents Pythia 1b with the average-case canaries, shows that using all the tokens is beneficial, however, it is not always the case. For instance, when evaluating the LSTM with average-case setting, \Cref{fig:group-lstm-average} shows that Min with a reduced sequence length of 2 gives the best MIA.

\end{document}